\documentclass[letterpaper, paper,11pt]{AAS} 

\usepackage{bm}

\usepackage{amsmath}

\usepackage{subcaption}

\usepackage{comment}

\usepackage[colorlinks=true, pdfstartview=FitV, linkcolor=black, citecolor= black, urlcolor= black]{hyperref}

\usepackage{footnpag}  

\usepackage{caption}

\usepackage{graphicx}

\usepackage{graphicx}

\usepackage{float}

\usepackage{amsmath}

\usepackage{amssymb}

\usepackage{bm} 

\usepackage{algorithm}

\usepackage{algpseudocode}

\usepackage{amsmath}

\usepackage{tikz}

\usepackage{caption}

\usepackage{float}

\usepackage{algorithm}

\usepackage{algpseudocode}

\usepackage{setspace}

\usetikzlibrary{arrows.meta, positioning, fit}

\usetikzlibrary{shapes.geometric, arrows.meta, positioning, fit, calc}

\usepackage{booktabs} 

\usepackage{multirow} 

\usepackage{tabularx}

\usepackage[backend=biber, style=ieee, doi=true, url=false]{biblatex}

\usepackage{hyperref}

\usepackage{booktabs}

\newcommand{\vect}[1]{\bm{#1}}

\newcommand{\mat}[1]{\bm{#1}}

\newcommand{\quat}[1]{\mathbf{#1}} 

\newcommand{\qmult}{\otimes} 

\PaperNumber{25-895}
\begin{document}

\title{Star Tracker Misalignment Compensation in Deep Space Navigation Through Model-Based Estimation}

\author{Ridma Ganganath \thanks{Ph.D. Student, Department of Aerospace Engineering, Iowa State, Ames, IA, USA.},  
\ Simone Servadio\thanks{Assistant Professor, Department of Aerospace Engineering, Iowa State University, Ames, IA, USA.},\ and David Lee\thanks{Assistant Professor, Department of Aerospace Engineering, Iowa State University, Ames, IA, USA.}
}

\maketitle

\begin{abstract}
This work presents a novel adaptive framework for simultaneously estimating spacecraft attitude and sensor misalignment. Uncorrected star tracker misalignment can introduce significant pointing errors that compromise mission objectives in GPS-denied environments. To address this challenge, the proposed architecture integrates a Bayesian Multiple-Model Adaptive Estimation (MMAE) framework operating over an $N \times N \times N$ 3D hypothesis grid. Each hypothesis employs a 9-state Multiplicative Extended Kalman Filter (MEKF) to estimate attitude, angular velocity, and gyroscope bias using TRIAD-based vector measurements. A key contribution is the development of a robust grid refinement strategy that uses hypothesis diversity and weighted-mean grid centering to prevent the premature convergence commonly encountered in classical, dominant model-based refinement triggers. Extensive Monte Carlo simulations demonstrate that the proposed method reduces the final misalignment RMSE relative to classical approaches, achieving arcsecond-level accuracy. The resulting framework offers a computationally tractable and statistically robust solution for in-flight calibration, enhancing the navigational autonomy of resource-constrained spacecraft.

\end{abstract}

\section{Introduction}

Accurate and autonomous spacecraft attitude determination is a cornerstone of deep-space CubeSat missions, where limited ground support and the absence of GPS demand entirely onboard solutions. Small satellites operating beyond Earth orbit cannot rely on facilities like NASA’s Deep Space Network, so they must achieve high-fidelity attitude knowledge despite uncertain dynamics, sensor imperfections, and environmental disturbances \cite{mckeen2025attitude1,servadio2020nonlinear,doi:10.2514/1.G004355}. In this context, star trackers and gyroscopes are often the primary sensors for attitude estimation on resource-constrained platforms. However, practical challenges such as small misalignments between a star tracker’s mounting frame and the spacecraft body frame can introduce persistent errors if not properly estimated and corrected. Ensuring robust attitude estimation under such conditions requires advanced filtering architectures that go beyond the classical single-model paradigm.

Traditional attitude filters like the Extended Kalman Filter (EKF) and its multiplicative variant (MEKF) have long been the industry standard due to their efficiency and ability to handle quaternion kinematics gracefully \cite{crassidis2007survey}. The MEKF, in particular, enforces the unit-quaternion constraint and avoids the singularities associated with Euler angles or other minimal attitude representations. While these filters perform well under nominal conditions, they typically assume a fixed sensor configuration and rely on a single model of the system. This single-model reliance can limit their robustness when confronted with unmodeled effects such as sensor misalignment, time-varying external disturbances, or non-Gaussian noise characteristics \cite{crassidis2004unscented}. In deep-space scenarios with minimal opportunities for ground recalibration, even a small star tracker misalignment can lead to a significant pointing error if the filter cannot adapt to it. Augmenting the state vector to estimate misalignment within a standard EKF/MEKF framework is possible but often poses convergence and observability issues, especially when the misalignment is constant and subtle.

To address these limitations, we propose a modular adaptive estimation framework that combines a 9-state MEKF with a Bayesian Multiple-Model Adaptive Estimation (MMAE) layer to jointly estimate the spacecraft’s attitude, gyroscope bias, angular velocity, and fixed star tracker misalignment. In the proposed architecture, the MEKF fuses gyroscope data and direction measurements, which are formulated as attitude observations via the TRIAD algorithm \cite{shuster1981three}. TRIAD provides a quaternion observation by fusing two non-collinear reference vectors (e.g., a sun vector and a star vector), offering a reliable attitude measurement source. This multi-model architecture avoids nonlinear state augmentation for misalignment estimation and supports parallel implementation, making it well-suited to CubeSat-class onboard processors. 

Our approach builds upon and extends several threads of prior research in adaptive filtering and spacecraft calibration. Earlier works have demonstrated the benefit of multiple-model methods in related contexts, such as adaptive calibration of star trackers and inertial sensors \cite{servadio2021differential}. For example, Lam et al. evaluated an MMAE scheme for improving on-orbit attitude determination accuracy by accounting for misalignment and other model uncertainties \cite{doi:10.2514/6.2007-6816}. Nebelecky and colleagues explored techniques for compensating model errors in attitude filters to enhance estimation fidelity under unknown disturbances \cite{6916281}. Crassidis et al. introduced a generalized MMAE framework that leverages residual autocorrelation to improve hypothesis discrimination in fault detection and calibration tasks \cite{4085937}. These studies underscore the value of model diversity and hypothesis testing in reliable spacecraft state estimation. Moreover, recent advances in uncertainty-aware filtering — such as the use of Koopman operator theory for nonlinear observer design under model uncertainty \cite{doi:10.2514/1.A35688} and particle filtering approaches that sample system dynamics from posterior distributions \cite{10643370} — motivate the inclusion of modular, robust inference architectures for complex systems. In contrast to prior methods, the framework presented here uniquely integrates a classical MEKF with a multi-hypothesis adaptation layer to handle sensor misalignments in real time, and introduces a quaternion-fusion step to maintain a high-confidence attitude solution drawn from multiple concurrent filters. 

In summary, this work provides a scalable and resilient attitude estimation solution for deep-space CubeSats facing sensor misalignment and other uncertainties. The proposed MEKF–MMAE architecture is capable of simultaneously estimating the spacecraft’s attitude and the true star tracker misalignment without requiring manual recalibration or state-vector augmentation. Monte Carlo simulations validate that the filter bank converges to the correct misalignment and consistently yields high-accuracy attitude knowledge, with errors remaining within $3\sigma$ uncertainty bounds. This adaptive, model-based approach thus enables high-fidelity autonomous navigation for resource-limited platforms, ensuring robust performance in GPS-denied environments where traditional single-model filters may falter.

\section{Problem Formulation}

We consider the spacecraft attitude estimation problem using a combination of two star-trackers and a gyroscope rigidly mounted on a CubeSat. The true attitude of the spacecraft is defined with respect to an inertial reference frame ( $\mathcal{I} = (x_I, y_I, z_I)$), fixed with respect to distant stars. The CubeSat body frame is denoted as $\mathcal{B} = (x_B, y_B, z_B)$ and is aligned with the spacecraft's structure. The first star tracker, ST$_1$, observes a known inertial vector to Star 1 ($\vect{v}_1^{I}$) and provides a corresponding measurement, $\vect{v}_1^{B}$, in its own local sensor frame. Similarly, the second star tracker, ST$_2$, observes a known inertial vector to Star 2 ($\vect{v}_2^{I}$) and provides a measurement, $\vect{v}_2^{B}$, in its local frame.

Importantly, due to the nonlinearity and randomness, there will be many imperfections in the system.  In this work, the objective is to estimate the spacecraft's attitude $\mathbf{q}_{BI}(t)$, angular velocity $\boldsymbol{\omega}(t)$, gyroscope bias $\mathbf{b}(t)$, and a small constant star tracker misalignment vector $\boldsymbol{\mu}_R$. The estimation framework combines a 9-state Multiplicative Extended Kalman Filter (MEKF) to track the attitude, angular velocity, and gyroscope bias, with a Multiple-Model Adaptive Estimation (MMAE) layer to infer the misalignment $\boldsymbol{\mu}_R$ from a discrete grid of candidate hypotheses. The star tracker misalignment introduces a fixed rotational offset between the body frame and the sensor frame, affecting the measured directions of known inertial-frame vectors. Noisy measurements of these vectors, along with gyroscope readings, are used to update the filter and refine both attitude and misalignment estimates. Figure~\ref{fig:frames} illustrates the relevant coordinate frames and sensor geometry involved in this joint estimation problem.

\begin{figure}[ht]
    \centering
    \includegraphics[width=0.5\linewidth]{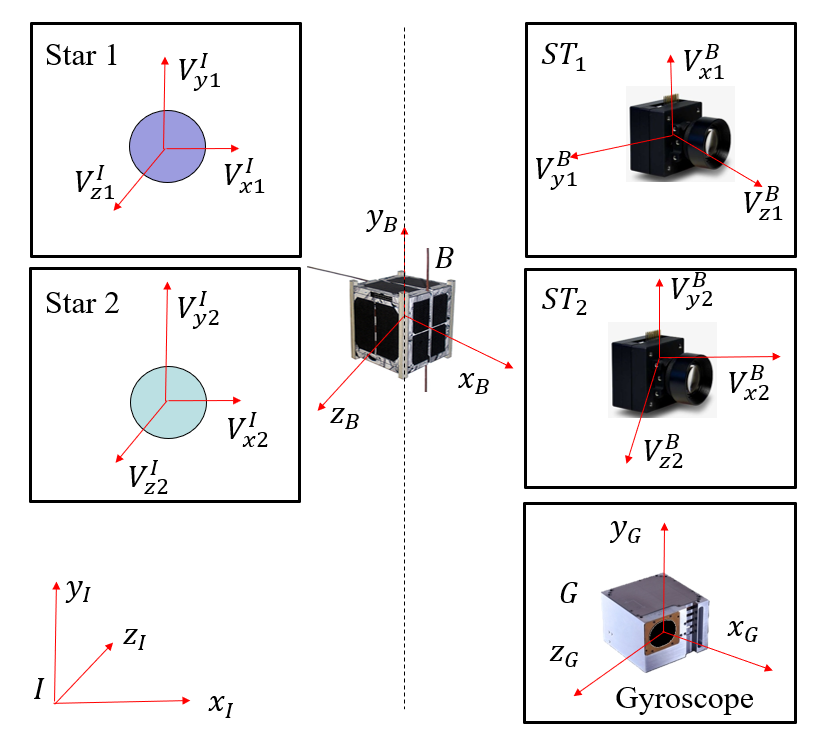} 
    \caption{Coordinate frames for the CubeSat, including the inertial ($\mathcal{I}$), body ($\mathcal{B}$), gyroscope (G), and two star-tracker ($ST_1$, $ST_2$) reference frames.}
    \label{fig:frames}
\end{figure}

\subsection{The Multiplicative Extended Kalman Filter (MEKF) with Star Tracker Measurement}

The MEKF serves as the foundational part of the architecture and is designed to estimate the spacecraft's attitude quaternion, angular velocity, and gyroscope bias in this problem. The filter operates on a multiplicative error state formulation to robustly handle the non-Euclidean nature of attitude quaternions. The true state of the system is defined as:
\begin{equation}
    \vect{x}_{\text{true}}(t) =
    \begin{bmatrix}
        \quat{q}(t) \\
        \vect{\omega}(t) \\
        \vect{b}(t)
    \end{bmatrix}
\end{equation}
where $\quat{q}(t) \in \mathbb{R}^4$ is the unit quaternion describing attitude (scalar-last convention), $\vect{\omega}(t) \in \mathbb{R}^3$ is the body-frame angular velocity, and $\vect{b}(t) \in \mathbb{R}^3$ represents the gyroscope bias.  The filter maintains a nominal state estimate, which can be described as:
\begin{equation}
\hat{\vect{x}}_{\text{nom}}(t) =
\begin{bmatrix}
\hat{\quat{q}}(t) \\
\hat{\vect{\omega}}(t) \\
\hat{\vect{b}}(t)
\end{bmatrix}
\end{equation}

The filter estimates a 9-dimensional error state, $\delta \vect{x}(t)$, which captures the deviation from the nominal estimate:
\begin{equation}
    \delta\vect{x}(t) =
    \begin{bmatrix}
        \delta\vect{\omega}(t) \\
        \delta\vect{b}(t) \\
        \delta\vect{\theta}(t)
    \end{bmatrix}
\end{equation}
where $\delta\vect{\omega}$ and $\delta\vect{b}$ are additive errors in the estimated angular velocity and bias, respectively. The attitude error is represented by a small-angle rotation vector, $\delta\vect{\theta} \in \mathbb{R}^3$, which relates the true and nominal quaternions via the multiplicative error quaternion described as:
\begin{equation}
    \delta\quat{q} \approx
    \begin{bmatrix}
        \frac{1}{2}\delta\vect{\theta} \\
        1
    \end{bmatrix}
\end{equation}
such that the corrected quaternion is evaluated via quaternion multiplication, can be defined as:
\begin{equation}
    \quat{q} = \delta\quat{q} \qmult \hat{\quat{q}}
\end{equation}

The uncertainty of this error state is captured by its covariance matrix is defined as:
\begin{equation}
    \mat{P}(t) = \mathbb{E}\left[\delta\vect{x}(t)\delta\vect{x}(t)^\top\right]
\end{equation}

\subsection*{System Dynamics}

The spacecraft’s true attitude is represented by a unit quaternion $\quat{q} \in \mathbb{R}^4$ (scalar-last convention) and the quaternion kinematics propagate as:
\begin{equation}
    \dot{\hat{\quat{q}}} = \frac{1}{2} \mat{\Omega}(\hat{\vect{\omega}}) \hat{\quat{q}} 
\end{equation}
where $\hat{\vect{\omega}} \in \mathbb{R}^3$ is the estimated body-frame angular velocity, and $\mat{\Omega}(\vect{\omega})$ is the quaternion multiplication matrix, which can be described as: 
\begin{equation}
    \mat{\Omega}(\vect{\omega}) =
    \begin{bmatrix}
        0 & -\omega_x & -\omega_y & -\omega_z \\
        \omega_x & 0 & \omega_z & -\omega_y \\
        \omega_y & -\omega_z & 0 & \omega_x \\
        \omega_z & \omega_y & -\omega_x & 0
    \end{bmatrix}
\end{equation}

In this formulation, the angular velocity evolves according to Euler’s rigid-body equations, with $M_c$ being the active external torques:
\begin{equation}
    \dot{\vect{\omega}} = \mat{J}^{-1} \left(M_c-\vect{\omega} \times (\mat{J}\vect{\omega}) \right)
\end{equation}
where $\mat{J}$ is the spacecraft's inertia matrix. The gyroscope bias $\vect{b} \in \mathbb{R}^3$ is modeled as constant:
\begin{equation}
    \dot{\vect{b}} = \mathbf{0}
\end{equation}

\subsection*{Prediction Step}
The MEKF is divided into the prediction step, when uncertainties are propagated under the dynamics, and a measurement update (or correction) step, where knowledge from the sensor is fused into the state uncertainties to obtain a more accurate estimate. 

The prediction step consists of two parallel operations. First, the nominal state $\hat{\vect{x}}_{\text{nom}}$ is propagated forward in time by numerically integrating the full nonlinear system dynamics. Simultaneously, the error-state covariance $\mat{P}$ is propagated using a linearized approximation based on the system Jacobians \cite{simon2006optimal}, which can be described as:
\begin{equation}
    \mat{P}_k^- = \mat{\Phi}_{k-1} \mat{P}_{k-1}^+ \mat{\Phi}_{k-1}^\top + \mat{Q} \Delta t
\end{equation}

where $\mat{P}_{k-1}^+$ denotes the a posteriori (updated) error covariance from the previous time step, and $\mat{P}_k^-$ represents the a priori (predicted) error covariance prior to incorporating the current measurement.
Here, $\mat{Q}$ denotes the continuous-time process noise covariance, and $\Delta t$ is the time increment between updates. The discrete-time state transition matrix $\mat{\Phi}$ assumes a block-diagonal structure:

\begin{equation}
    \mat{\Phi} =
    \begin{bmatrix}
        \exp(\mat{F} \Delta t) & \mathbf{0} & \mathbf{0} \\
        \mathbf{0} & \mat{I} & \mathbf{0} \\
        \mathbf{0} & \mathbf{0} & \mat{I}
    \end{bmatrix}
\end{equation}

where $\mat{F}$ is the Jacobian matrix of the angular velocity dynamics, and it can be described as:
\begin{equation}
    \mat{F} = -\mat{J}^{-1} \left( [\mat{J}\hat{\vect{\omega}}]_\times - [\hat{\vect{\omega}}]_\times \mat{J} \right)
\end{equation}
such that $[\cdot]_\times$ denotes the skew-symmetric operator that maps a vector to its cross-product matrix.

\subsection*{TRIAD-Based Attitude Measurement Model}

At each measurement update, the spacecraft attitude is estimated using the TRIAD algorithm \cite{doi:10.2514/3.2555}, which computes the optimal rotation aligning two inertial-frame reference vectors with their corresponding measurements in the body frame. Let $\vect{v}_1^I$ and $\vect{v}_2^I$ be known reference vectors expressed in the inertial frame, and let $\vect{v}_1^B$ and $\vect{v}_2^B$ denote their noisy observations in the body frame. From these, two orthonormal TRIAD bases are constructed and can be described as:
\begin{align*}
    \vect{t}_1^I &= \frac{\vect{v}_1^I}{\|\vect{v}_1^I\|} & 
    \vect{t}_1^B &= \frac{\vect{v}_1^B}{\|\vect{v}_1^B\|} \\
    \vect{t}_2^I &= \frac{\vect{v}_1^I \times \vect{v}_2^I}{\|\vect{v}_1^I \times \vect{v}_2^I\|} &
    \vect{t}_2^B &= \frac{\vect{v}_1^B \times \vect{v}_2^B}{\|\vect{v}_1^B \times \vect{v}_2^B\|}\\
    \vect{t}_3^I &= \vect{t}_1^I \times \vect{t}_2^I &
    \vect{t}_3^B &= \vect{t}_1^B \times \vect{t}_2^B
\end{align*}

These vectors define the inertial-frame and body-frame TRIAD matrices and can be demonstrated as:
\begin{equation}
    \mat{T}_I = \begin{bmatrix} \vect{t}_1^I & \vect{t}_2^I & \vect{t}_3^I \end{bmatrix}, \quad
    \mat{T}_B = \begin{bmatrix} \vect{t}_1^B & \vect{t}_2^B & \vect{t}_3^B \end{bmatrix}
\end{equation}
The estimated direction cosine matrix (DCM) from inertial to body frame is computed as:
\begin{equation}
    \hat{\mat{C}}_{BI} = \mat{T}_B \mat{T}_I^\top
\end{equation}
which is then converted into a quaternion representation ($\quat{q}_{\text{meas}}$), and serves as the TRIAD-based attitude measurement input for the filter update.

\subsection*{Update Step}

The measurement update step incorporates the TRIAD-derived quaternion and gyroscope measurements to correct the nominal state. The attitude residual quaternion is then computed as:
\begin{equation}
    \delta\quat{q}_{\text{res}} = \quat{q}_{\text{meas}} \qmult \hat{\quat{q}}^*
\end{equation}
where $\hat{\quat{q}}^*$ denotes the conjugate of the nominal quaternion estimate $\hat{\quat{q}}$. Assuming small attitude errors, this residual is converted into a Modified Rodrigues Parameter (MRP) vector as:
\begin{equation}
    \vect{s}_{\text{res}} = \text{q2mrp}(\delta\quat{q}_{\text{res}})
\end{equation}

The angular velocity residual is defined as the difference between the measured angular velocity and the predicted value\cite{KOTTATH201688}:
\begin{equation}
\vect{\omega}_{\text{res}} = \vect{\omega}_{\text{meas}} - (\hat{\vect{\omega}} + \hat{\vect{b}})
\end{equation}
where $\hat{\vect{b}}$ is the estimated gyroscope bias. The residual of the composite measurement becomes:
\begin{equation}
    \vect{y} =
    \begin{bmatrix}
        \vect{s}_{\text{res}} \\
        \vect{\omega}_{\text{res}}
    \end{bmatrix}, \quad
    \mat{H} =
    \begin{bmatrix}
        \mathbf{0} & \mathbf{0} & \mat{I} \\
        \mat{I} & \mat{I} & \mathbf{0}
    \end{bmatrix}
\end{equation}
where $\mat{H}$ is the linearized measurement Jacobian.  The innovation covariance $\mat{S}$ and Kalman gain $\mat{K}$ are computed as:
\begin{gather}
    \mat{S} = \mat{H} \mat{P}^- \mat{H}^\top + \mat{R} \\
    \mat{K} = \mat{P}^- \mat{H}^\top \mat{S}^{-1} \\
    \delta\vect{x}^+ = \mat{K} \vect{y} \\
    \mat{P}^+ = (\mat{I} - \mat{K} \mat{H}) \mat{P}^- (\mat{I} - \mat{K} \mat{H})^\top + \mat{K} \mat{R} \mat{K}^\top
\end{gather}
where $\mat{P}^-$ is the prior error covariance, and $\mat{R}$ is the measurement noise covariance. The attitude estimate is then corrected using the small-angle approximation, where the attitude error component $\delta\vect{\theta}$ is extracted from $\delta\vect{x}^+$. This vector is converted into a quaternion correction and described as:
\begin{align}
    \delta\quat{q} &= \text{mrp2q}(\delta\vect{\theta}) \\
    \hat{\quat{q}}^+ &= \delta\quat{q} \qmult \hat{\quat{q}}^-, \quad \hat{\quat{q}}^+ \leftarrow \frac{\hat{\quat{q}}^+}{\|\hat{\quat{q}}^+\|}
\end{align}

where rotational correction, $\delta\vect{\theta}$, is extracted from $\delta\vect{x}^+$, while the nominal angular velocity, $\hat{\vect{\omega}}$, and bias, $\hat{\vect{b}}$, are updated with their corresponding components. Finally, the error state is reset to zero, which completes the measurement update cycle.

\subsection*{Measurement Noise Models for TRIAD Observations}
To assess the MEKF's sensitivity to sensor noise, we consider two prevalent noise models applied to the body-frame inputs of the TRIAD algorithm.

\textbf{Additive Noise Model:} The additive model perturbs each rotated inertial vector with zero-mean Gaussian noise and can be described as:
\begin{align}
    \vect{v}_i^B &= \mat{C}_{BI} \vect{v}_i^I + \vect{\eta}_i 
\end{align}
where $\mat{C}_{BI}$ would be the true direction cosine matrix (DCM) from inertial to body frame, $\vect{v}_i^I$ are known inertial vectors ($i = 1,2$), and $\vect{\eta}_i \sim \mathcal{N}(\vect{0}, \sigma_v^2 \mat{I}_3)$ is zero-mean Gaussian vector noise.

\textbf{Multiplicative Noise Model:} In contrast, the multiplicative model introduces uncertainty through small random rotations, and can be stated as:
\begin{align}
    \vect{v}_i^B &= \mat{C}_{\delta_i} \mat{C}_{BI} \vect{v}_i^I
\end{align}

where $\mat{C}_{\eta_i} = \text{mrp2dcm}(\delta\vect{\theta}_i)$ is a small random rotation matrix generated from zero-mean attitude noise, $\delta\vect{\theta}_i \sim \mathcal{N}(\vect{0}, \sigma_\theta^2 \mat{I}_3)$.

Although both models are evaluated, the multiplicative model is used throughout this work due to its higher fidelity in simulating star tracker behavior. As shown in the Figure~\ref{fig:additive_vs_multiplicative} and Table~\ref{tab:noise_comparison}, it yields more accurate and consistent TRIAD-based attitude estimates under realistic conditions.

\subsection{The Robust Multiple-Model Adaptive Estimation (MMAE) Framework for Joint Attitude and Star Tracker Misalignment Estimation }

This approach augments the classical 9-state Multiplicative Extended Kalman Filter (MEKF) by estimating the star tracker misalignment vector ($\boldsymbol{\mu}_R$) using a grid-based Multiple Model Adaptive Estimation (MMAE) framework. Attitude is estimated using the TRIAD method, and the optimal quaternion is computed via Markley quaternion fusion. Each model in the MMAE bank corresponds to a hypothesis of $\boldsymbol{\mu}_R$.

\subsection*{Hypothesis Grid}

Let $\bm{y}_k \in \mathbb{R}^m$ denote the measurement at time step $k$, such as a TRIAD-based attitude quaternion converted to a Modified Rodrigues Parameters (MRP) vector. Assume a bank of $N$ competing models $\mathcal{M}_j$, each corresponding to a distinct hypothesis of the star tracker misalignment vector $\bm{\mu}_R^{(j)} \in \mathbb{R}^3$. Let $\hat{\bm{y}}_k^{(j)}$ denote the predicted measurement under model $\mathcal{M}_j$, $\bm{r}_j = \bm{y}_k - \hat{\bm{y}}_k^{(j)}$ be the residual under model $j$, $\bm{R} \in \mathbb{R}^{m \times m}$ the measurement noise covariance matrix, and $w_j^{(k-1)}$ the prior probability (weight) of model $\mathcal{M}_j$ at time $k-1$.

We aim to compute the posterior model probability given a new measurement $\bm{y}_k$ via a Bayesian approach and can be defined as:
\begin{equation}
P(\mathcal{M}_j \mid \bm{y}_k) = \frac{ P(\bm{y}_k \mid \mathcal{M}_j) P(\mathcal{M}_j) }{ P(\bm{y}_k) }
\end{equation}
Using the model prior weight, which is its probability after prediction, $P(\mathcal{M}_j) = w_j^{(k-1)}$, and marginal likelihood $P(\bm{y}_k) = \sum_{h=1}^N P(\bm{y}_k \mid \mathcal{M}_h) w_h^{(k-1)}$, we get the weight update equation as:
\begin{equation}
w_j^{(k)} =
\frac{
w_j^{(k-1)} \cdot P(\bm{y}_k \mid \mathcal{M}_j)
}{
\sum\limits_{h=1}^{N} w_h^{(k-1)} \cdot P(\bm{y}_k \mid \mathcal{M}_h)
}
\end{equation}

Assuming additive zero-mean Gaussian measurement noise in the measurement model, regardless of how the true measurement is affected by the stochastic variable, The likelihood of measurement $\bm{y}_k$ under model $\mathcal{M}_j$ is:
\begin{equation}
    P(\bm{y}_k \mid \mathcal{M}_j) =
    \frac{1}{(2\pi)^{m/2} \sqrt{\det \bm{R}}} \exp \left( -\frac{1}{2} \bm{r}_j^\top \bm{R}^{-1} \bm{r}_j \right)
\end{equation}

Substituting into the Bayesian weight update then gives as:
\begin{equation}
    w_j^{(k)} =
    \frac{
    w_j^{(k-1)} \cdot \exp \left( -\frac{1}{2} \bm{r}_j^\top \bm{R}^{-1} \bm{r}_j \right)
    }{
    \sum\limits_{h=1}^{N} w_h^{(k-1)} \cdot \exp \left( -\frac{1}{2} \bm{r}_h^\top \bm{R}^{-1} \bm{r}_h \right)
    }
\end{equation}
which provides the Maximum Likelihood-based Bayesian model probability update.

\subsection*{Implementation of Specific Residual Definition}

The residual for each model is the rotational error between the TRIAD-based measurement quaternion ($\bm{q}_{\text{meas}}$) and the quaternion predicted by that model ($\bm{q}_j$). This error is expressed as a 3D small-angle vector in the Modified Rodrigues Parameters (MRP) domain, and can be described as:

\begin{equation}
\vect{s}_{\text{res}}^{(j)} = \text{MRP}\left( \bm{q}_{\text{meas}} \otimes \hat{\bm{q}}_j^{*} \right)
\end{equation}

To maintain numerical stability, the weight update is performed in two steps. First, an un-normalized weight is computed by multiplying the prior weight by the likelihood of the residual, which is assumed to be Gaussian, which can be stated as:
\begin{equation}
\tilde{w}_j^{(k)} = w_j^{(k-1)} \cdot \exp\left(-\frac{1}{2} \bm{r}_j^\top \bm{R}^{-1} \bm{r}_j \right)
\end{equation}
This intermediate value incorporates the new measurement information. The final weight is then found by normalizing across all models to ensure the weights sum to one, can be described as:
\begin{equation}
w_j^{(k)} =
\frac{
\tilde{w}_j^{(k)}
}{
\sum\limits_{h=1}^{N} \tilde{w}_h^{(k)}
}
\end{equation}
This two-step process provides a robust, recursive update rule for the model probabilities.

\subsection*{Pruning: Eliminating Low-Weight Models}

Over time, many hypotheses become irrelevant (i.e., have low weight). To improve computational efficiency, a pruning threshold, $w_\text{prune}$, is introduced to eliminate them. First, the set of all valid hypotheses is identified as those whose weights exceed this threshold as:
\begin{equation}
\mathcal{J}_\text{valid}^{(k)} = \left\{ j \mid w_j(k) > w_\text{prune} \right\}
\end{equation}
This step effectively creates a list of all models that remain consistent with the measurement data. The filter bank is then reduced, retaining only these statistically significant hypotheses:
\begin{equation}
\left\{ \bm{\mu}_R^{(j)} \right\}_{j=1}^N \rightarrow \left\{ \bm{\mu}_R^{(j)} \right\}_{j \in \mathcal{J}_\text{valid}^{(k)}}
\end{equation}
Finally, the weights of the all surviving models are renormalized to ensure their sum remains one, preserving a valid probability distribution, can be desceibed as:

\begin{equation}
w_j(k) \leftarrow \frac{ w_j(k) }{ \sum\limits_{h \in \mathcal{J}_\text{valid}^{(k)}} w_h(k) }, \quad \forall j \in \mathcal{J}_\text{valid}^{(k)}
\end{equation}

This complete pruning process reduces computational complexity in subsequent updates by focusing resources only on the most probable models.



\subsection*{Grid Branching and Refinement Strategy}

Estimating the sensor misalignment presents a conflict between precision and computational cost. A globally high-resolution hypothesis grid is computationally infeasible, while a coarse grid is inaccurate. Adaptive refinement resolves this by starting coarse and iteratively ``zooming in" on the most probable solution. This strategy achieves a high-precision estimate by focusing computational resources only wher it needed, making the problem tractable and efficient.

The trigger for this refinement is a crucial design choice. In this work, we investigate and compare three distinct triggering mechanisms to determine the most robust strategy.

\begin{enumerate}
    \item \textbf{Classical Trigger with Dominant Model-Based Centering:} The common and classical approach is to refine when a single model becomes dominant, identified by its weight exceeding a predefined threshold. 

    Let \( w_j^{(k)} \) denote the weight of the \( j \)-th hypothesis \( \boldsymbol{\mu}_R^{(j)} \) at time step \( k \). The first trigger condition is met when the maximum model weight exceeds a predefined branching threshold \( w_{\text{branch}} \), can be described as:
\begin{equation}
\max_j w_j^{(k)} > w_{\text{branch}}
\label{eq:branch_threshold}
\end{equation}
Once the condition (Equation (\ref{eq:branch_threshold})) is satisfied, the Maximum A Posteriori (MAP) index \( j^* \) and the corresponding misalignment hypothesis are found:
\begin{equation}
j^* = \arg\max_j w_j^{(k)}, \quad \hat{\boldsymbol{\mu}}_R^{\text{MAP}} = \boldsymbol{\mu}_R^{(j^*)}
\label{eq:map_estimate}
\end{equation}

We then generate a locally refined hypothesis grid around \( \hat{\boldsymbol{\mu}}_R^{\text{MAP}} \). Upon refinement, all model states are reset to the state of the MAP hypothesis, and the weights are reinitialized to be uniform.

\item \textbf{Hypothesis Diversity \(\Psi(t)\) Trigger with Dominant Model-Based Centering:} In this work, we propose a more robust dual-trigger strategy. In addition to the maximum weight criterion, we monitor the overall hypothesis diversity using the $\Psi$ metric \cite{10643370}, which is analogous to the effective number of particles in particle filters. To quantify model diversity over time, we define the weight uniformity metric \(\Psi(t)\) as:
\begin{equation}
    \Psi(t) = 100 \cdot \frac{A_t}{N}
    \label{eq:psi}
\end{equation}

\begin{equation}
    \qquad A_t = \left( \sum_{i=1}^N u_i^2(t) \right)^{-1}
    \label{eq:psi}
\end{equation}
where \(N\) is the number of models in the current hypothesis set and \(u_i(t)\) is the normalized weight of model \(i\) at time \(t\). The term \(A_t\) represents the inverse of the weight concentration (also known as the effective number of models). A value of \(\Psi(t) = 100\%\) implies a perfectly uniform model distribution, while low $\Psi$ value indicates that the filter has successfully localized the solution, even if no single model is dominant.  When the refinement is triggered by a low $\Psi$ value, the new grid is centered on the hypothesis with the Maximum A Posteriori (MAP) probability at that instant. This prevents the filter from getting stuck in a state of local consensus where no single model is confident enough to trigger a refinement on its own, but a direction of likely vectors is highlighted.

\item \textbf{Hypothesis Diversity \(\Psi(t)\) Trigger with Weighted Mean-Based Centering:} This final strategy builds upon the diversity-based trigger but improves the grid recentering mechanism. While the trigger condition remains the same ( 10\%\ $\Psi(t)$ value), the new refined grid is centered on the weighted mean of the hypothesis set rather than the MAP estimate, which is described in Equation \eqref{eq:mean_estimate}, described as:

\begin{equation}
\hat{\boldsymbol{\mu}}_R^{\text{Mean}} = \sum_{j=1}^{N} w_j^{(k)} \boldsymbol{\mu}_R^{(j)}
\label{eq:mean_estimate}
\end{equation}

\end{enumerate}


\vspace{1em}



\subsection*{Optimal Attitude Fusion via Quaternion Averaging}

At each time step, the MMAE filter produces a bank of \( N \) distinct attitude quaternions, \( \hat{\quat{q}}^{(j)} \), each with an associated posterior probability, \( w^{(j)} \). The final step is to fuse this distributed information into a single, optimal attitude estimate that best represents the entire set of hypotheses. For this, we employ the statistically optimal quaternion averaging method developed by Landis Markley \cite{averageq}.

The method seeks the mean quaternion, \( \hat{\quat{q}} \), that minimizes the weighted sum of the squared chordal distances to each hypothesis quaternion in the bank, can be explained as:
\begin{equation}
\hat{\quat{q}} = \arg\min_{\quat{q} \in \mathbb{S}^3} \sum_{j=1}^{N} w^{(j)} \left\| \quat{q} - \hat{\quat{q}}^{(j)} \right\|^2
\label{eq:min_chordal}
\end{equation}

The literature demonstrated that this optimization problem can be elegantly solved by constructing a \( 4 \times 4 \) symmetric matrix, \( \mat{M} \), which aggregates the weighted outer products of all attitude hypotheses, which can be described as:

\begin{equation}
\mat{M} = \sum_{j=1}^{N} w^{(j)} \hat{\quat{q}}^{(j)} \left( \hat{\quat{q}}^{(j)} \right)^T
\label{eq:markley_matrix}
\end{equation}

Then the optimal fused quaternion, \( \hat{\quat{q}} \), that satisfies the minimization criterion in Equation~\eqref{eq:min_chordal} is the eigenvector of \( \mat{M} \) corresponding to its maximum eigenvalue, given as:
\begin{equation}
\hat{\quat{q}} = \mathrm{eigvec}_{\max}(\mat{M})
\label{eq:markley_fused_q}
\end{equation}

To ensure the result remains a valid unit quaternion on the hypersphere \( \mathbb{S}^3 \), a final normalization is performed. Furthermore, to address the inherent sign ambiguity of the quaternion representation (\( \quat{q} \equiv -\quat{q} \)), the resulting quaternion is aligned to maintain consistency with the previous time step’s estimate. This fusion process, combined with the adaptive grid refinement, forms the core of a robust MEKF-MMAE framework capable of accurate joint estimation.

\section{Numerical Simulations}

Our entire problem in this work was validated and simulated using non-linear rigid-body dynamics, as described in the problem formulation. The simulation initializes the spacecraft with a constant angular velocity vector of [3.0, 4.4, -5.0] deg/s, emulating a steady rotation phase,  to avoid unobservable scenarios. To replicate realistic post-maneuver behavior, a damping torque is applied after $t = 4100$\,s. This braking torque is modeled as $\boldsymbol{\tau}_\text{damp} = -D \boldsymbol{\omega}$, where $D = 0.6$ is a scalar damping coefficient. The full angular velocity dynamics thus become:
\begin{equation}
\dot{\boldsymbol{\omega}} = \mathbf{J}^{-1} \left( \boldsymbol{M}_c - \boldsymbol{\omega} \times (\mathbf{J} \boldsymbol{\omega}) - D \boldsymbol{\omega} \right)
\quad
\label{manu}
\end{equation}
where $t\geq 4100$, causing the spacecraft's spin rate to decay toward zero gradually. The simulations were conducted over 100 Monte Carlo runs, each lasting 5000 seconds with a 0.5~s time-step, for a spacecraft with an inertia matrix of 
$\bm{J} = \mathrm{diag}(100,\ 60,\ 50)\ \mathrm{kg} \cdot \mathrm{m}^2$. 
The filter's tuning parameters, including the initial covariance ($\bm{P}_0$), process noise ($\bm{Q}$), and measurement noise ($\bm{R}$), are detailed in Table~\ref{tab:covariance_matrices}. 
For the adaptive estimation, the process began with a $7 \times 7 \times 7$ 3D hypothesis grid and used a hypothesis diversity ($\Psi$) threshold of 10\% to trigger the refinement.

\subsection{ Multiple-Model Adaptive Estimation  Simulation   }

The proposed framework, illustrated in Figure~\ref{fig:frames}, employs an MMAE architecture to jointly estimate spacecraft attitude and star tracker misalignment. The system utilizes a bank of MEKFs, where each filter is conditioned on a distinct misalignment hypothesis, $\boldsymbol{\mu}_R^{(j)}$.

\begin{figure}[ht]
    \centering
    \includegraphics[width=0.8\linewidth]{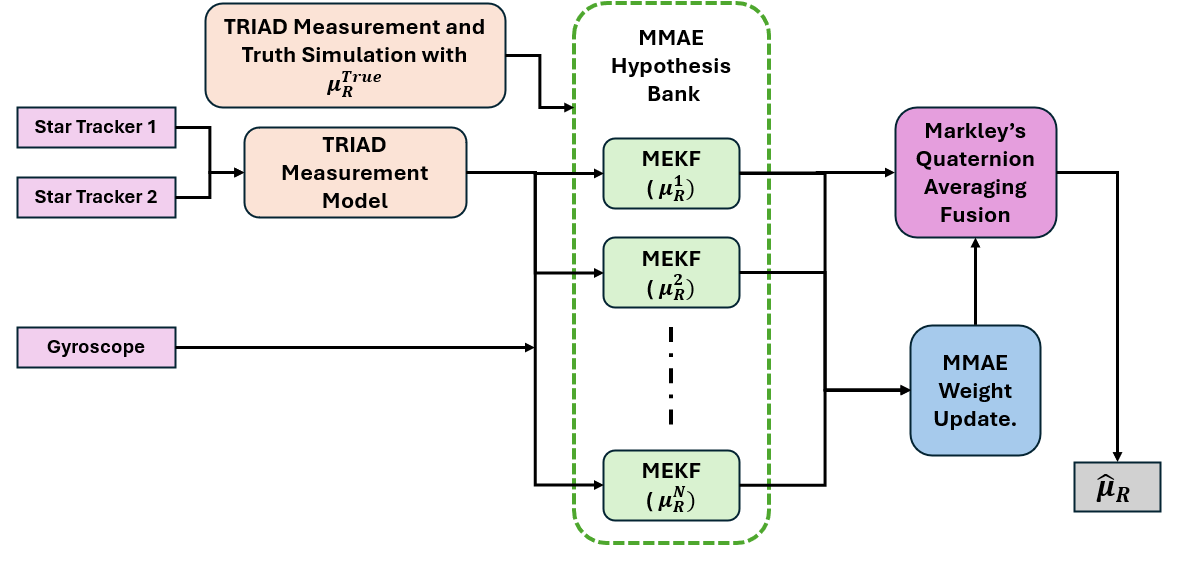} 
    \caption{MMAE Architecture for Attitude Determination. }
    \label{fig:frames}
\end{figure}

All filters process shared measurements derived from the TRIAD algorithm and gyroscope data. At each timestep, the measurement residuals from each MEKF are used to compute a likelihood, which updates the posterior probability (weight) of the corresponding hypothesis. The Maximum A Posteriori (MAP) hypothesis guides the adaptive grid refinement, while the final misalignment estimate is computed as the weighted mean of all active hypotheses. A robust attitude estimate is obtained by fusing the outputs of all individual MEKFs using Markley's weighted quaternion averaging. The core logic of this process is summarized in Algorithm~\ref{alg:mmae_mekf}.

\begin{algorithm}[H]
\caption{The MMAE Algorithm for Attitude and Misalignment Estimation}
\label{alg:mmae_mekf}
\begin{algorithmic}[1]
\State \textbf{Initialize:} Nominal state $(\hat{\quat{q}}_0, \hat{\boldsymbol{\omega}}_0, \hat{\boldsymbol{b}}_0)$ and covariance $\mat{P}_0$
\State \textbf{Hypotheses:} Generate a small-angle misalignment grid $\{\boldsymbol{\mu}_R^{(j)}\}_{j=1}^N$ and assign uniform weights $w_j = 1/N$
\State \textbf{Filter Bank:} Initialize $N$ MEKF instances, each corresponding to a hypothesis $\boldsymbol{\mu}_R^{(j)}$
\For{each timestep $k$}
    \State Propagate truth state and generate measurements $(\quat{q}_{\text{meas}}, \boldsymbol{\omega}_{\text{meas}})$
    \For{each model $j$}
        \State Predict MEKF state $(\hat{\quat{q}}^{(j)}, \hat{\boldsymbol{\omega}}^{(j)}, \hat{\boldsymbol{b}}^{(j)})$
        \State Form expected measurement $\hat{\quat{q}}_{\text{exp}}^{(j)} = \quat{q}_{\mu}^{(j)} \otimes \hat{\quat{q}}^{(j)}$
        \State Compute residual $\mathbf{y}_k^{(j)} = [\text{MRP}(\quat{q}_{\text{meas}} \otimes \hat{\quat{q}}_{\text{exp}}^{(j),-1}); \ \boldsymbol{\omega}_{\text{meas}} - (\hat{\boldsymbol{\omega}}^{(j)} + \hat{\boldsymbol{b}}^{(j)})]$
        \State Update MEKF state using residual $\mathbf{y}_k^{(j)}$
        \State Update weight: $w_j \gets w_j \cdot L(\mathbf{y}_k^{(j)})$
    \EndFor
    \State Normalize weights: $w_j \gets w_j / \sum_{l=1}^N w_l$
    \State Compute diversity metric: $\Psi_k = \frac{100}{N \sum_j w_j^2}$
    \If{$\Psi_k < \Psi_{\text{th}}$ and refinements remain}
        \State Compute weighted mean: $\hat{\boldsymbol{\mu}}_R = \sum_j w_j \boldsymbol{\mu}_R^{(j)}$
        \State Generate refined hypothesis grid centered at $\hat{\boldsymbol{\mu}}_R$
        \State Propagate new models from highest-weighted (MAP) filter 
        \State Reset weights to uniform  $w_j = 1/N$
    \Else
        \State  Prune low-weight models and renormalize weights
    \EndIf
    \State Fuse attitude via optimal quaternion averaging
    \State Store state estimates and misalignment error
\EndFor
\end{algorithmic}
\end{algorithm}

\section{Results and Analysis}

\subsection{Additive Vs. Multiplicative Results and Analysis}

\begin{figure}[h]
    \centering
    \includegraphics[width=0.6\textwidth]{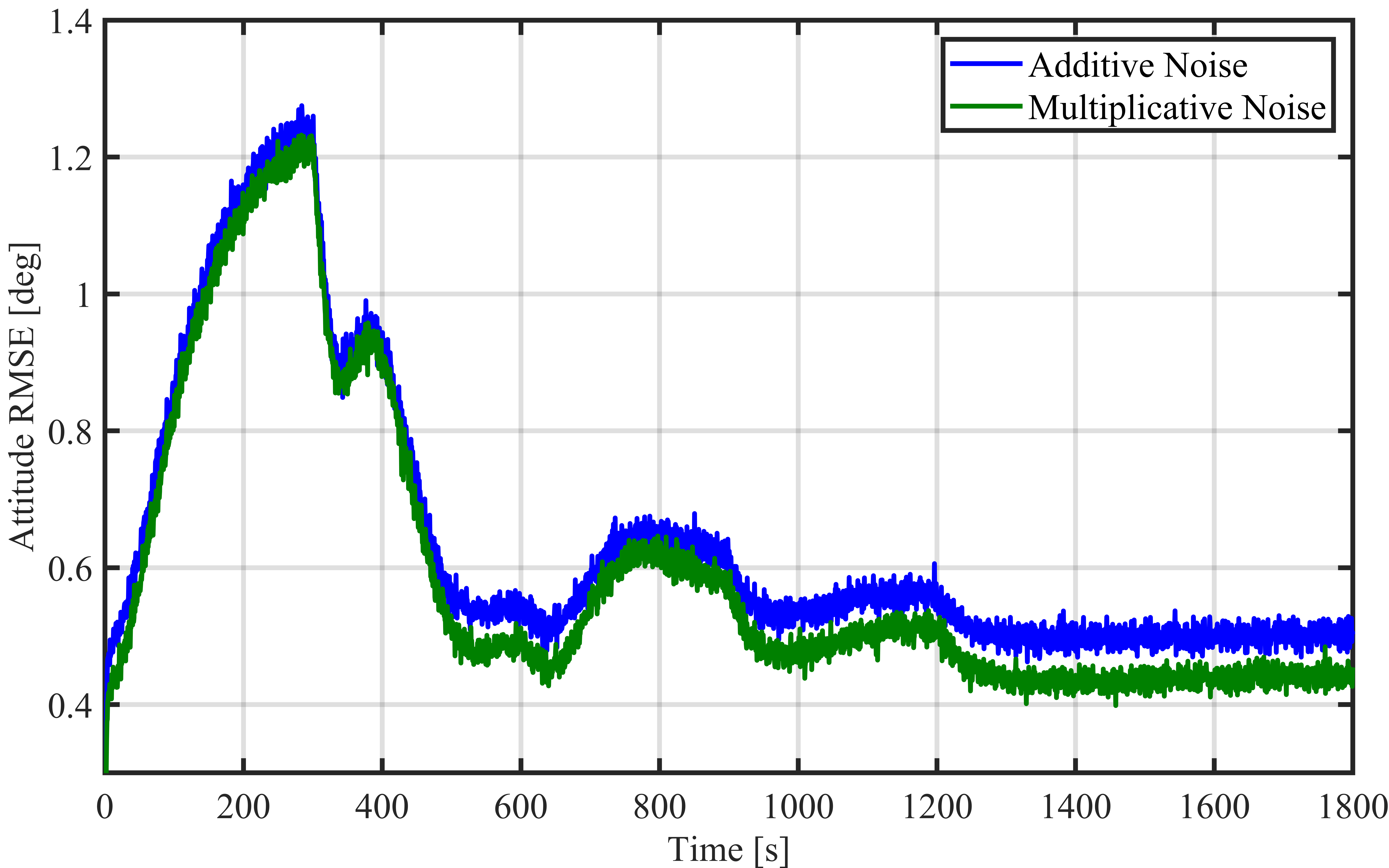}
    \caption{Attitude RMSE Comparison: Additive vs. Multiplicative TRIAD Noise}
    \label{fig:additive_vs_multiplicative}
\end{figure}

The MEKF's performance was evaluated under both additive and multiplicative TRIAD noise models to determine the most physically representative formulation for star tracker error. The results of this comparison, averaged over 100 Monte Carlo runs, are presented in Figure~\ref{fig:additive_vs_multiplicative} and quantified in Table~\ref{tab:noise_comparison}. 

\begin{table}[h]
    \centering
    \caption{Final Attitude Error Comparison: Additive vs. Multiplicative TRIAD Noise}
    \label{tab:noise_comparison}
    \begin{tabular}{@{}lcc@{}} 
        \toprule
        \textbf{Metric} & \textbf{Additive} & \textbf{Multiplicative} \\
        \midrule
        Mean Final Error (deg) & 0.4640 & 0.3887 \\
        Std Final Error (deg)  & 0.1915 & 0.1663 \\
        Max Final Error (deg)  & 1.1207 & 1.1217 \\
        \bottomrule
    \end{tabular}
\end{table}

To ensure the filter was tested in a realistic scenario, a series of maneuvers were simulated in this specific MEKF example, with step changes in the spacecraft's true angular velocity occurring at 5, 10, 15, and 20 minutes into the simulation. Under these dynamic conditions, the multiplicative model consistently outperforms the additive model in attitude estimation accuracy. Specifically, it achieves a 16\% lower mean final error (0.3887$^\circ$ vs. 0.4640$^\circ$) and a reduced final standard deviation (0.1663$^\circ$ vs. 0.1915$^\circ$). While the maximum final error remains comparable (1.1217$^\circ$ vs. 1.1207$^\circ$), these results highlight the higher accuracy of the multiplicative formulation in capturing the inherently rotational nature of sensor pointing errors. This is due to the nonlinear transformation of the noise stochastic variable.

\begin{table}[h]
\centering
\caption{Summary of estimation errors at simulation end ($t=5000$~s) over 100 Monte Carlo runs.}
\label{tab:mekf_summary}
\begin{tabular}{@{} l l c @{}} 
\toprule
\textbf{Error Metric} & \textbf{Statistic} & \textbf{Value} \\
\midrule
\multirow{3}{*}{Attitude Error (deg)} & Mean & 0.4511 \\
 & Std. Dev. & 0.1644 \\
 & Max & 0.7868 \\
\midrule
\multirow{2}{*}{Angular Velocity Error (rad/s)} & Mean & 7e-05 \\
 & Std. Dev. & 3e-05 \\
\midrule
\multirow{2}{*}{Gyroscope Bias Error (rad/s)} & Mean & 0.00012 \\
 & Std. Dev. & 5e-05 \\

\bottomrule
\end{tabular}
\end{table}

These results validate the TRIAD-enhanced MEKF as a reliable and high-precision estimator. This robust performance is critical, as this filter serves as the computational backbone for each hypothesis within the MMAE framework used to address the sensor misalignment problem.

\subsection*{Multiple-Model Adaptive Estimation Results and Analysis }

The performance of the proposed MMAE framework was evaluated through 100 Monte Carlo simulations to ensure statistical robustness. Each run was initialized with the same filter parameters, including the initial covariance matrix $\mat{P}_0$, process noise covariance $\mat{Q}$, and measurement noise covariance $\mat{R}$, as summarized in Table~\ref{tab:covariance_matrices}.

\begin{table}[h]
\centering
\caption{Covariance and Noise Matrix Values}
\label{tab:covariance_matrices}
\begin{tabular}{@{} l l c @{}} 
\toprule
\textbf{Matrix} & \textbf{Component} & \textbf{Value Used} \\
\midrule
\multirow{3}{*}{\bm{$P_0$}} & Angular Velocity & $(0.01)^2 \ \mathrm{rad}^2/\mathrm{s}^2$ \\
 & Gyroscope Bias & $(0.001)^2 \ \mathrm{rad}^2/\mathrm{s}^2$ \\
 & Attitude Error (MRP) & $(1.0)^2 \ \mathrm{rad}^2$ \\
\midrule
\multirow{3}{*}{\bm{$Q$}} & Angular Velocity & $(1 \times 10^{-6})^2 \ \mathrm{rad}^2/\mathrm{s}^4$ \\
 & Gyroscope Bias Drift & $(5 \times 10^{-8})^2 \ \mathrm{rad}^2/\mathrm{s}^4$ \\
 & Attitude Drift & $(5 \times 10^{-7})^2 \ \mathrm{rad}^2$ \\
\midrule
\multirow{2}{*}{\bm{$R$}} & Star Tracker (Attitude) & $(8.73 \times 10^{-4})^2 \ \mathrm{rad}^2$ \\
 & Gyroscope (Rate) & $(0.0005)^2 \ \mathrm{rad}^2/\mathrm{s}^2$ \\
\bottomrule
\end{tabular}
\end{table}

To provide a comprehensive measure of the filter's performance, the Root Mean Square Error (RMSE) \cite{servadio2025dynamicalupdatemapsparticle} for attitude ($\Xi_{q,k}$), angular velocity ($\Xi_{\omega,k}$), gyro bias ($\Xi_{b,k}$), and misalignment ($\Xi_{\mu,k}$) are calculated using equations \eqref{eq:rmse_q} through \eqref{eq:rmse_mu}.
\begin{align}
    \Xi_{q,k} &= \sqrt{\frac{1}{N_{MC}} \sum_{i=1}^{N_{MC}} \left( \mathbf{q}_{T,k}^{(i)} - \hat{\mathbf{q}}_{k}^{(i)+} \right)^T \left( \mathbf{q}_{T,k}^{(i)} - \hat{\mathbf{q}}_{k}^{(i)+} \right)} \label{eq:rmse_q} \\
    \Xi_{\omega,k} &= \sqrt{\frac{1}{N_{MC}} \sum_{i=1}^{N_{MC}} \left( \boldsymbol{\omega}_{T,k}^{(i)} - \hat{\boldsymbol{\omega}}_{k}^{(i)+} \right)^T \left( \boldsymbol{\omega}_{T,k}^{(i)} - \hat{\boldsymbol{\omega}}_{k}^{(i)+} \right)} \label{eq:rmse_w} \\
    \Xi_{b,k} &= \sqrt{\frac{1}{N_{MC}} \sum_{i=1}^{N_{MC}} \left( \mathbf{b}_{T,k}^{(i)} - \hat{\mathbf{b}}_{k}^{(i)+} \right)^T \left( \mathbf{b}_{T,k}^{(i)} - \hat{\mathbf{b}}_{k}^{(i)+} \right)} \label{eq:rmse_b} \\
    \Xi_{\mu,k} &= \sqrt{\frac{1}{N_{MC}} \sum_{i=1}^{N_{MC}} \left( \boldsymbol{\mu}_{T,k}^{(i)} - \hat{\boldsymbol{\mu}}_{k}^{(i)+} \right)^T \left( \boldsymbol{\mu}_{T,k}^{(i)} - \hat{\boldsymbol{\mu}}_{k}^{(i)+} \right)} \label{eq:rmse_mu}
\end{align}

\begin{figure}[h]
    \centering
    \includegraphics[width=4.5in,height=4.5in,keepaspectratio]{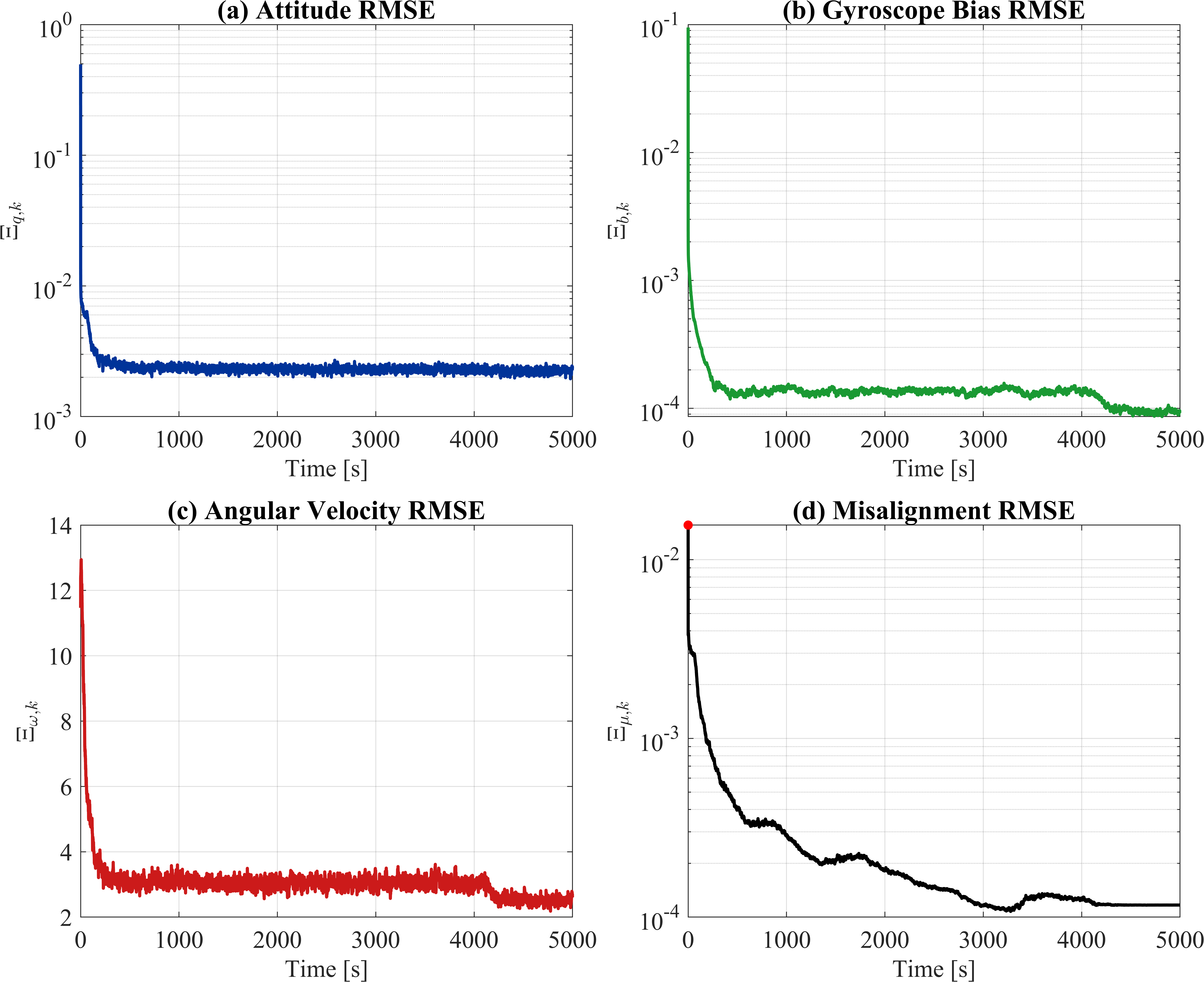}

    \caption{Root Mean Square Error (RMSE) for each estimated state, averaged over 100 Monte Carlo runs. The plots show the convergence of the errors for (a) attitude, (b) gyroscope bias, (c) angular velocity, and (d) star tracker misalignment.
    }
    \label{fig:refinement_psi_dual_trigger}
\end{figure}

The general convergence of the filter is quantified in Figure \ref{fig:refinement_psi_dual_trigger}. Each subplot demonstrates that the estimation error for its respective state rapidly decreases from the initial uncertainty and settles at a stable, low-error value.  The attitude RMSE~(a) settles to approximately $2 \times 10^{-3}$, while the star tracker misalignment RMSE~(d) converges to $1.5 \times 10^{-4}$, validating the framework's ability to identify the unknown sensor error. The gyroscope bias~(b) and angular velocity~(c) errors also showed a robust convergence towards low values.

Table~\ref{tab:mmae_results} provides a final snapshot of the component-wise mean errors at the end of the simulation horizon ($t=5000$~s), further confirming the high-fidelity performance of the filter. The results confirm arc-second accuracy, with final attitude and misalignment errors on the order of hundredths of a degree per axis. Furthermore, the final angular velocity and gyroscope bias errors are approximately $10^{-5}$~rad/s and $10^{-6}$~rad/s, respectively, demonstrating the micro-radian-level precision required for demanding deep-space applications.

\begin{table}[h]
\small
\captionsetup{width=\textwidth}
\centering
\caption{Multiple-Model Adaptive Estimation: Final Mean Errors at $t=5000.0$~s over 100 Monte Carlo runs.}
\begin{tabular}{@{} p{3.2cm} c r @{\hskip 0.5cm} p{3.2cm} c r @{}}
\toprule
\textbf{Error Metric} & \textbf{Axis} & \textbf{Mean Error} & \textbf{Error Metric} & \textbf{Axis} & \textbf{Mean Error} \\
\midrule
\label{tab:mmae_results}
\multirow{3}{=}{Attitude Error (deg)} 
 & X & 0.007101 & \multirow{3}{=}{Angular Velocity Error (rad/s)} 
 & X & $1.237 \times 10^{-5}$ \\
 & Y & 0.006030 &  & Y & $-2.93 \times 10^{-6}$ \\
 & Z & -0.021401 &  & Z & $-1.237 \times 10^{-5}$ \\
\midrule
\multirow{3}{=}{Gyroscope Bias Error (rad/s)} 
 & X & $3.6 \times 10^{-6}$ & \multirow{3}{=}{Misalignment Error (deg)} 
 & X & -0.001213 \\
 & Y & $6.98 \times 10^{-6}$ &  & Y & 0.000469 \\
 & Z & $-2.4 \times 10^{-7}$ &  & Z & 0.001374 \\
\bottomrule
\end{tabular}
\end{table}

\begin{figure}[H]
    \centering

    \begin{minipage}[t]{0.49\textwidth}
        \centering
        \textbf{(a)} Attitude component estimation errors \\
        \includegraphics[width=\linewidth]{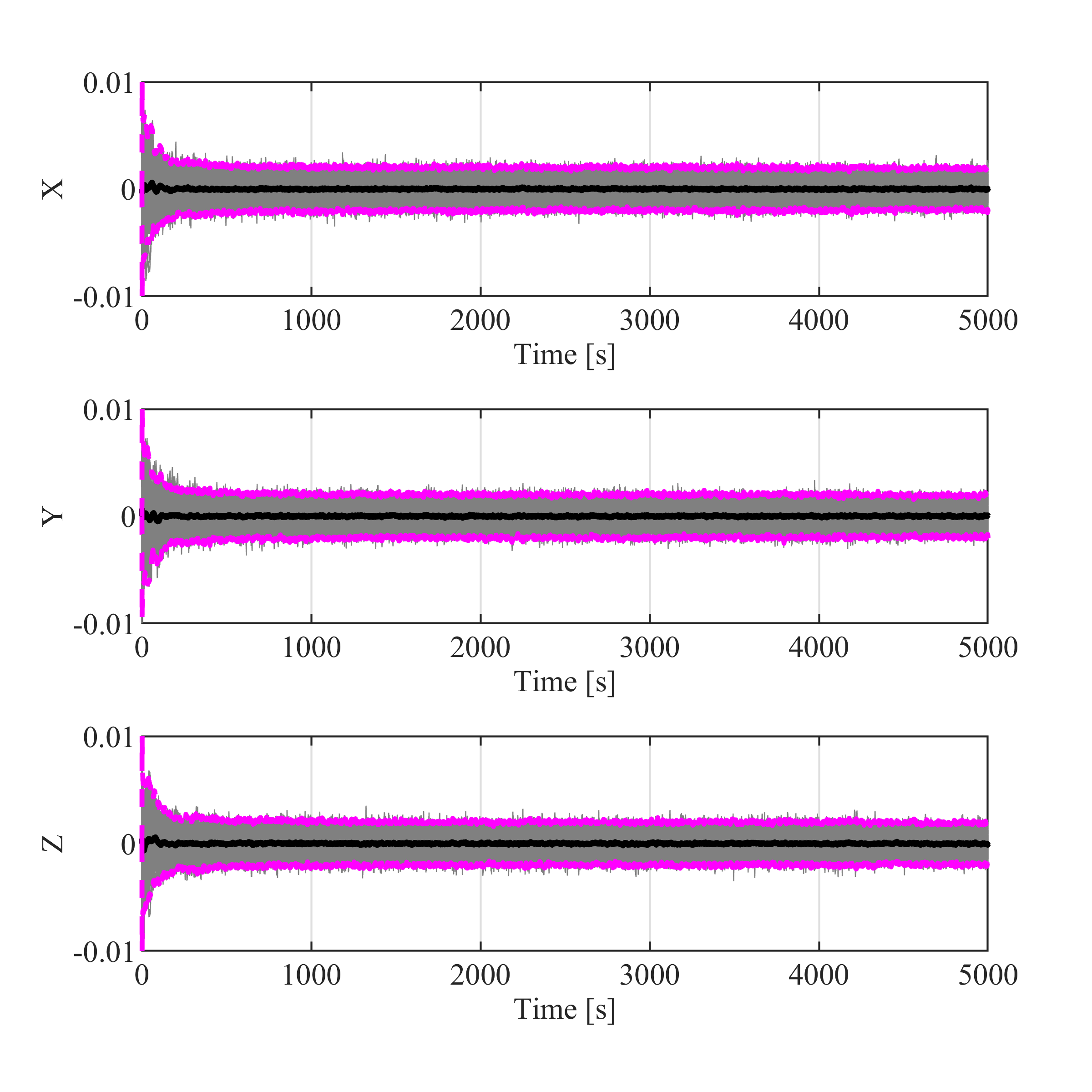}
    \end{minipage}
    \hfill
    \begin{minipage}[t]{0.482\textwidth}
        \centering
        \textbf{(b)} Gyroscope bias estimation errors \\
        \includegraphics[width=\linewidth]{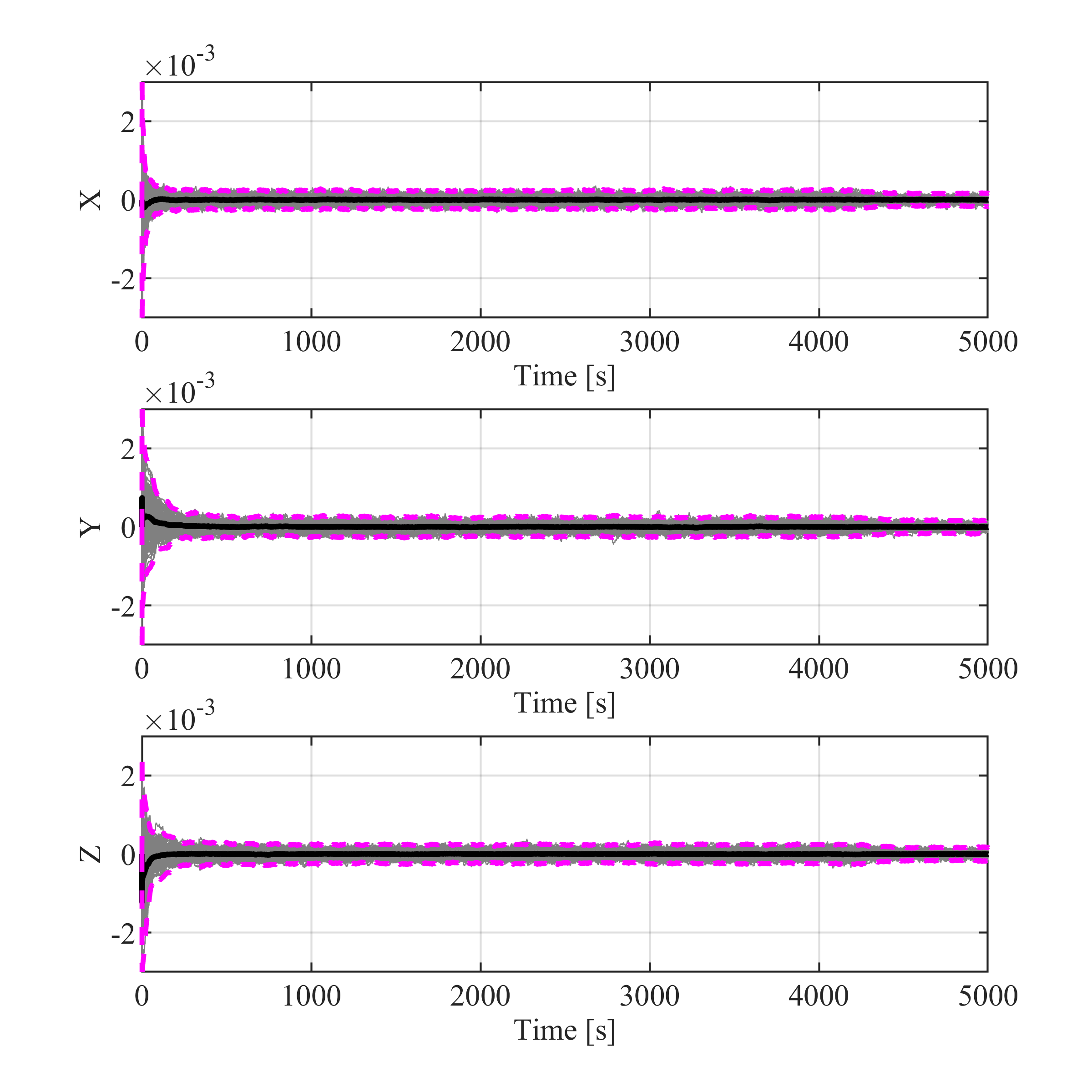}
    \end{minipage}

    \vspace{4mm}

    \begin{minipage}[t]{0.49\textwidth}
        \centering
        \textbf{(c)} Angular velocity estimation errors \\
        \includegraphics[width=\linewidth]{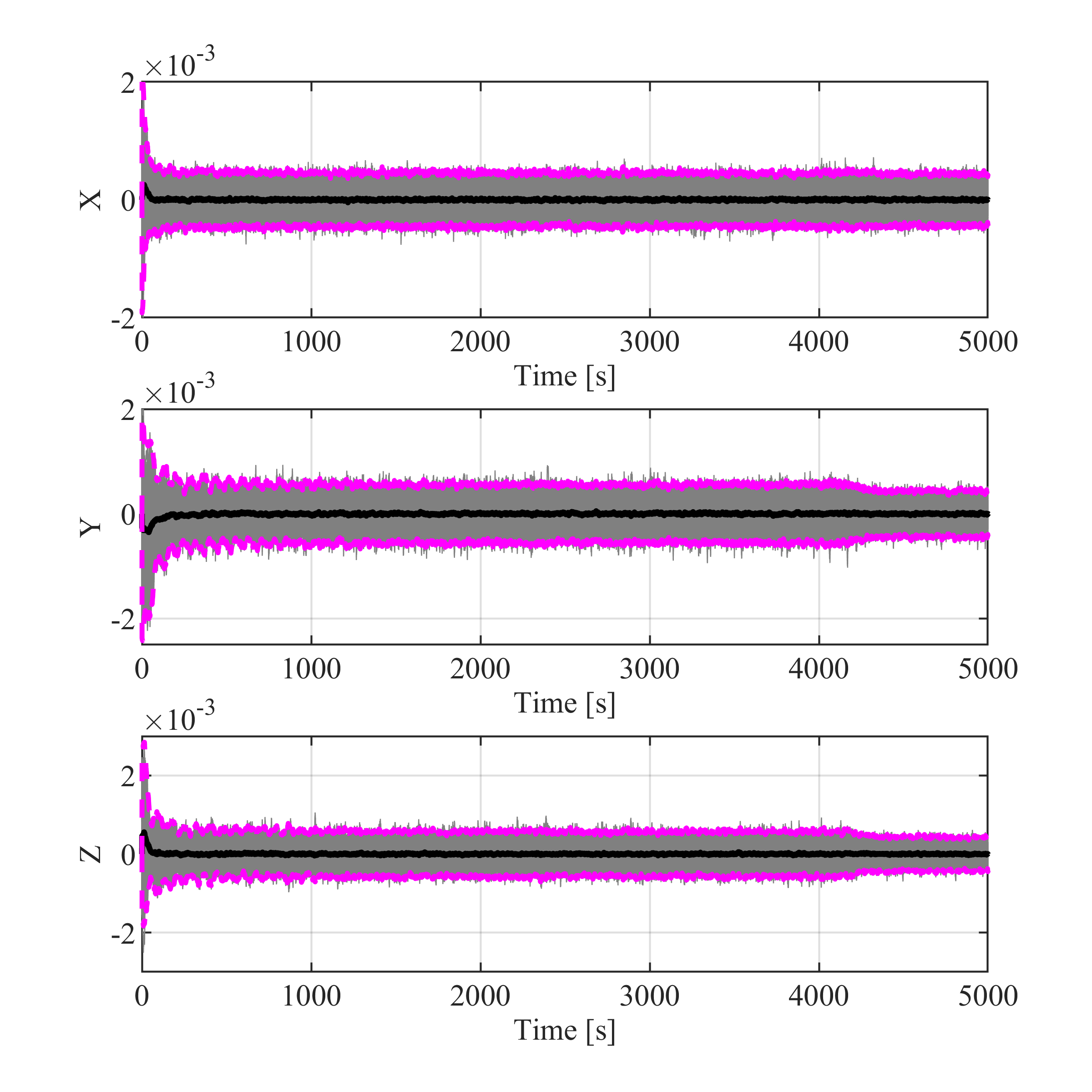}
    \end{minipage}
    \hfill
    \begin{minipage}[t]{0.488\textwidth}
        \centering
        \textbf{(d)} Star tracker misalignment estimation errors \\
        \includegraphics[width=\linewidth]{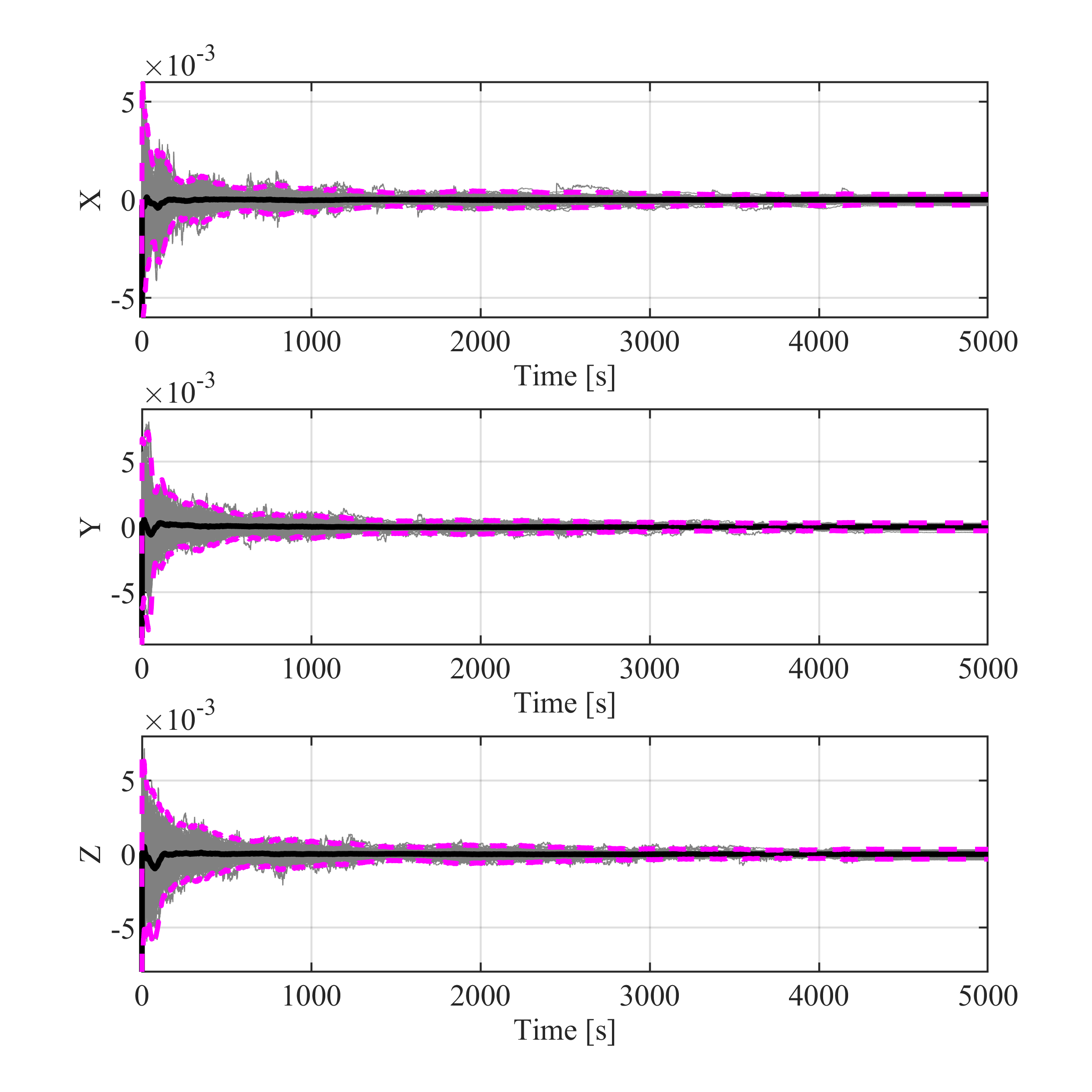}
    \end{minipage}

    \caption{
        Estimation errors over time from Monte Carlo dimulations. 
        (a) Attitude estimation errors using quaternion components ($q_x$, $q_y$, $q_z$). 
        (b) Gyroscope bias estimation errors ($b_x$, $b_y$, $b_z$). 
        (c) Angular velocity estimation errors ($\omega_x$, $\omega_y$, $\omega_z$). 
        (d) Star-Tracker misalignment estimation errors ($\mu_{R,x}$, $\mu_{R,y}$, $\mu_{R,z}$). 
        All plots include $\pm3\sigma$ bounds across 100 Monte Carlo trials.
    }
    \label{fig:4panel_errors}
\end{figure}

For further validation, a filter consistency analysis was also performed. The filter's ability to jointly estimate the spacecraft attitude and sensor misalignment is demonstrated and reported in Figure~\ref{fig:4panel_errors}, which presents the time evolution of the estimation errors for all state components. The plots illustrate that the errors for attitude, gyroscope bias, angular velocity, and star tracker misalignment all converge rapidly and remain within the statistically expected $\pm3\sigma$ covariance bounds. This confirms that the filter is well tuned and provides a consistent and reliable state estimate.

\subsection*{Comparative Analysis of Adaptive Refinement Strategies}

\begin{figure}[H]
    \centering
    \includegraphics[width=1\textwidth]{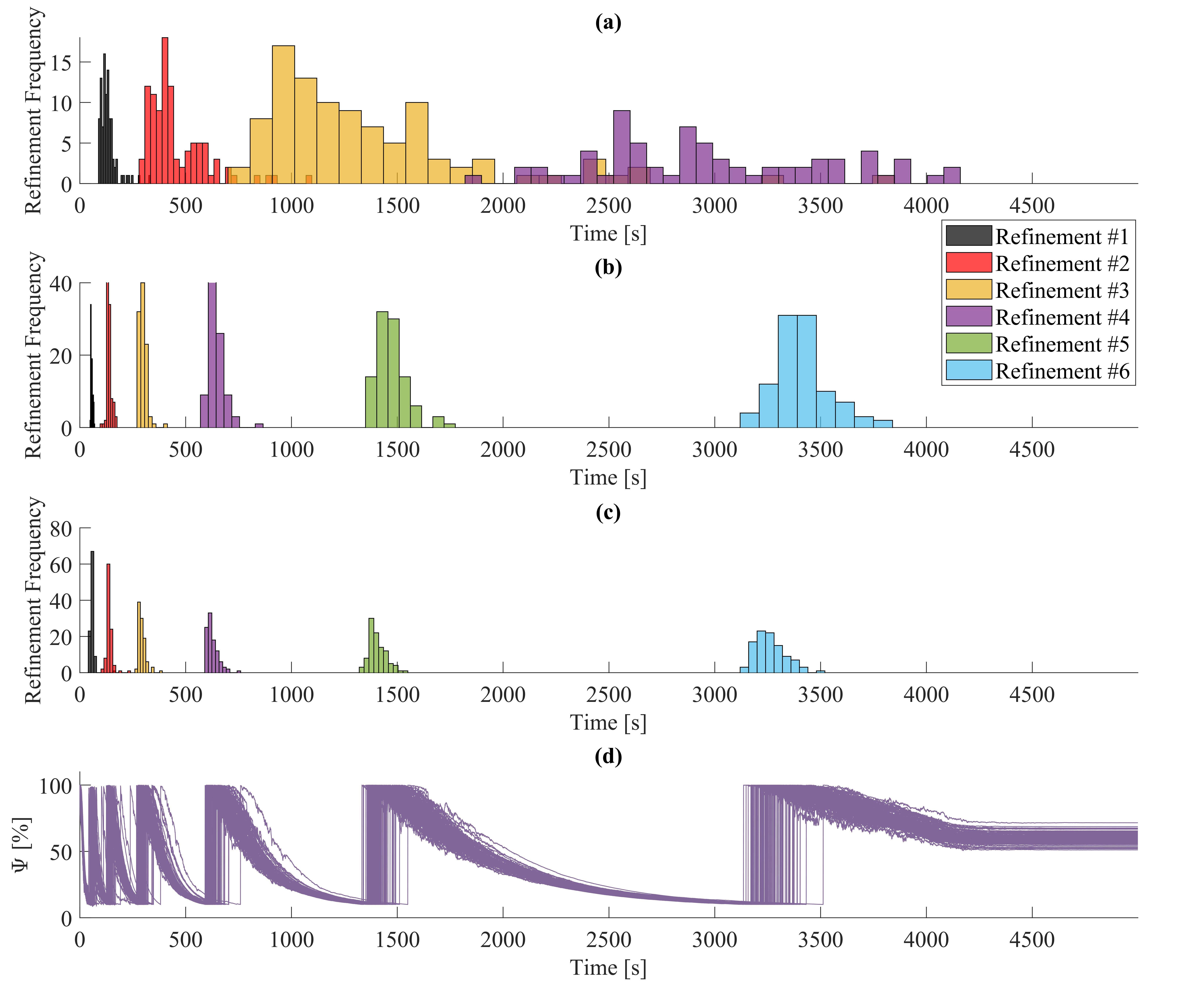}
   \caption{%
        Comparison of adaptive refinement strategies and resulting hypothesis diversity.
        \textbf{(a):} Classical Trigger (MAP Center). 
        \textbf{(b):} $\Psi$-Based Trigger (MAP Center). 
        \textbf{(c):} $\Psi$-Based Trigger (mean Center). 
        \textbf{(d):} Evolution of the Hypothesis Diversity Metric, $\Psi$, over 100 Monte Carlo runs for the $\Psi$-Based Trigger (mean Center) case.
    }
    \label{fig:refinement_comparison} 
\end{figure}

The behavior of the dual-trigger adaptive refinement strategy is illustrated in Figure~\ref{fig:refinement_comparison}. The upper subplot presents a histogram of the refinement event times across 100 Monte Carlo runs, while the lower subplot displays the corresponding evolution of the Hypothesis Diversity Metric, $\Psi$, corresponding to the strategy shown in Figure~\ref{fig:refinement_comparison}(d). The $\Psi$ based methodology continues refining until the grid of hypotheses reaches measurement noise levels, where it becomes impossible for the MMAE to prefer a few models over the other. The refinement mechanics stop when the MMAE reaches its steady state, shown by the constant values around 50\% to 80\% of models being effectively active, and therefore providing no preferred refinement direction. 

The classical weight-based trigger is the least effective strategy due to premature convergence, stalling at a final RMSE of \( 1.999 \times 10^{-4} \) after only 3.69 average refinements. This failure occurs because the trigger requires a single model to exceed a 50\% weight threshold. Late in the simulation, while overall model diversity (\( \Psi \)) becomes low due to a cluster of good hypotheses, no single model becomes dominant enough to trigger the refinement threshold. Therefore, the trigger mechanism traps the filter in a sub-optimal state, preventing the final refinements needed to eliminate remaining errors.

\begin{figure}[H]
    \centering
    \includegraphics[width=0.9\textwidth]{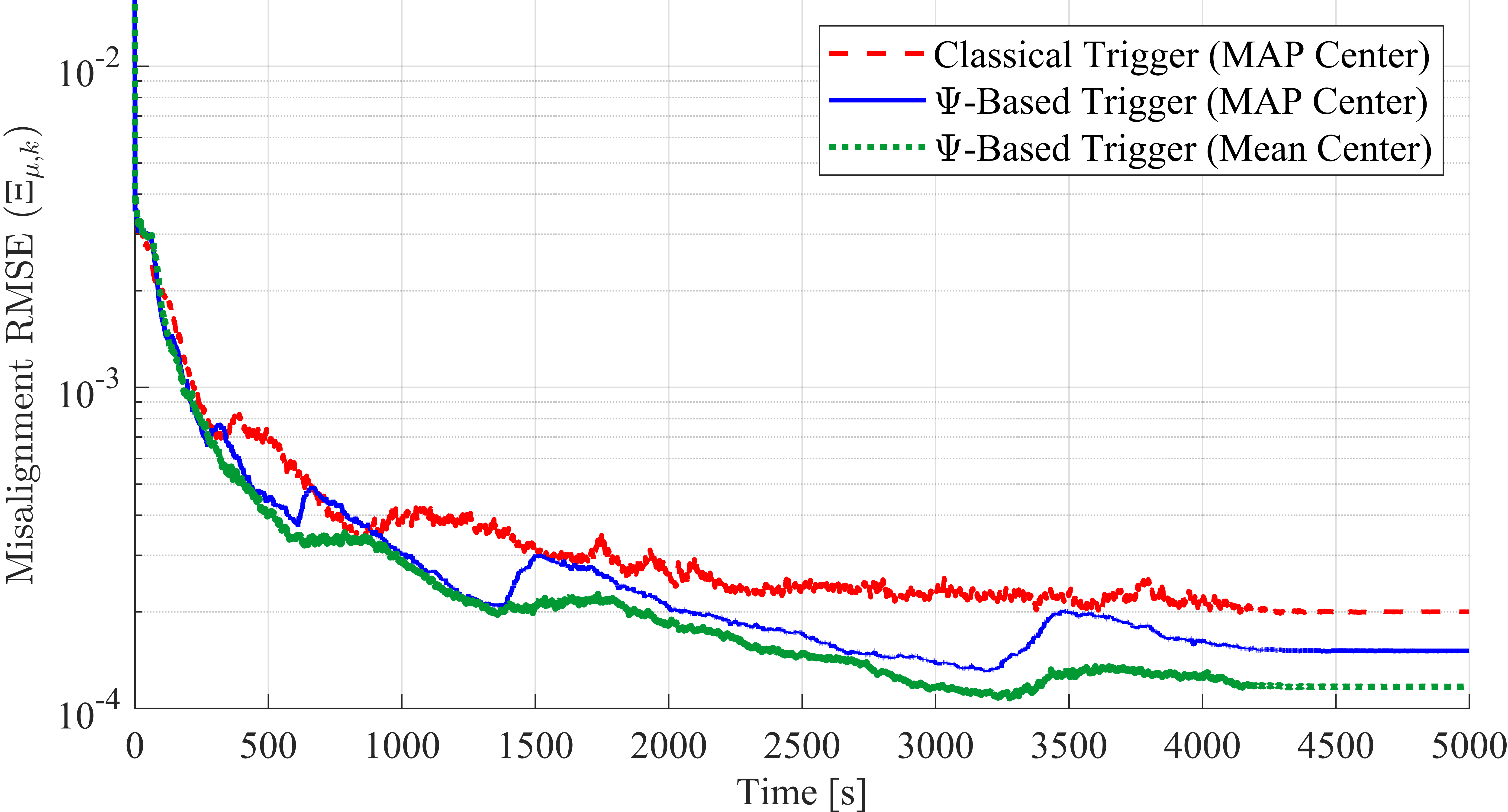}
    \caption{Comparison of misalignment RMSE ($\Xi_{\mu,k}$) between the proposed $\Psi(t)$-based trigger and the classical weight-based trigger. The proposed method achieves a 41.57\% reduction in final error by initiating additional refinements where the classical method stalls.}
    \label{fig:rmse_comparison}
\end{figure}

The advantage of the $\Psi$-based triggers is their ability to overcome the premature convergence of the classical method, as shown in Figure~\ref{fig:rmse_comparison}. These triggers initiate further refinements late in the simulation (near $t=1500$~s and $t=3300$~s), leading to continued error reduction. Although both $\Psi$-based strategies average the same number of refinements, the weighted mean centering approach is demonstrably superior. In contrast to the MAP-based strategy, which uses the highest-weighted hypothesis at that instant for grid centering, the mean-centering approach leverages the full posterior distribution by calculating a weighted average of all models. This method is more robust and provides a better estimate of the high-probability region's center, resulting in more effective refinements and the lowest final misalignment RMSE of  $1.168 \times 10^{-4}$ radians ($\approx 24.08$ arcseconds).

The analysis confirms that the proposed $\Psi$-based trigger with weighted mean centering provides the most accurate and robust performance. By preventing premature convergence and leveraging the full posterior distribution for grid refinement, this strategy reduces the final misalignment RMSE by 24.42\% compared to the $\Psi$-based MAP-centered approach and by a significant 41.57\% compared to the classical triggering approach. This demonstrates the method's robustness and confirms its selection as the optimal approach for this high-fidelity estimation problem.

\begin{table}[H]
\centering
\caption{Final Star-Tracker Misalignment RMSE ($\Xi_{\mu,k}$) and Average Refinement Counts for all Strategies over 100 Monte Carlo runs. }
\label{tab:avg_refinements}
\begin{tabular}{@{} l r r @{}} 
\toprule
\textbf{Strategy} & \textbf{Average Refinements} & \textbf{Final RMSE ($\Xi_{\mu,k}$)}(radians) \\
\midrule
Classical Trigger (MAP Center)              & 3.69 & $1.999 \times 10^{-4}$ \\
$\Psi$-Based Trigger (MAP Center)           & 6.00 & $1.511 \times 10^{-4}$ \\
$\Psi$-Based Trigger (Weighted Mean Center) & 6.00 & $1.168 \times 10^{-4}$ \\
\bottomrule
\end{tabular}
\end{table}

\section{Conclusion}

This paper introduced and validated a novel, model-based adaptive framework for the simultaneous estimation of spacecraft attitude and star-tracker misalignment in GPS-denied deep-space environments. The proposed architecture successfully integrates a bank of Multiplicative Extended Kalman Filters within a Multiple-Model Adaptive Estimation scheme, leveraging a newly introduced refinement strategy based on hypothesis diversity ($\Psi$) to achieve high-fidelity results without state-vector augmentation. A key contribution of this work would be the demonstrated accuracy of the proposed refinement strategy over classical and weight-based triggers. The analysis revealed that classical triggers are not prone to premature convergence, stalling when faced with a cluster of high-probability hypotheses. In contrast, the $\Psi$-based trigger correctly identifies this filter consensus and continues to refine the grid. The optimal performance was achieved with a weighted-mean centering approach, which leverages the full posterior distribution at each step. This method resulted in a 42\% reduction in the final misalignment Root Mean Square Error compared to the classical approach, confirming its superior accuracy and robustness.   

The significance of these findings lies in the direct applicability to autonomous, resource-constrained CubeSat missions. By providing a computationally tractable method for real-time, in-flight sensor calibration, this framework enhances the reliability and precision of deep-space navigation systems. Future work focuses on extending this framework to simultaneously estimate separate misalignment vectors, one for each instrument in a dual-star-tracker configuration, to further increase spacecraft autonomy.

\section*{Acknowledgments}
The authors wish to acknowledge the support of this work through the National Aeronautics and Space Administration (NASA) Established Program to Stimulate Competitive Research (EPSCoR) grant number 80NSSC24M0110.

\printbibliography

\end{document}